\documentclass[aps,prc,superscriptaddress,preprint]{revtex4-1}

\usepackage{amsmath,amssymb,bm}
\usepackage{graphicx,graphics,color}
\usepackage[colorlinks=true,linkcolor=blue,bookmarks=false]{hyperref}
\usepackage{epsfig}
\usepackage{latexsym}
\usepackage{rotating}
\usepackage{subfigure}
\usepackage{centernot}
\usepackage{longtable}
\usepackage{hhline,multirow,tabularx}
\usepackage{float}
\usepackage{mathrsfs}
\usepackage{xr}
\usepackage{slashed}
\usepackage{adjustbox}

\newcommand{\CL}{\mathcal{L}}
\newcommand{\CO}{\mathcal{O}}

\newcommand{\qv}{\vec{q}}
\newcommand{\sv}{\vec{S}}
\newcommand{\vv}{\vec{v}}
\newcommand{\pv}{\vec{p}}
\newcommand{\kv}{\vec{k}}
\newcommand{\xv}{\vec{x}}
\newcommand{\sigmav}{\vec{\sigma}}
\newcommand{\lv}{\vec{l}}
\newcommand{\gv}{\vec{\gamma}}

\begin{document}

\preprint{ADP-20-36/T1146}

\title{Relativistic mean-field corrections for interactions of dark matter particles with nucleons}

\author{X. G. Wang}
\affiliation{ARC Centre of Excellence for Dark Matter Particle Physics and CSSM, Department of Physics, University of Adelaide, SA 5005, Australia}
\author{A. W. Thomas}
\affiliation{ARC Centre of Excellence for Dark Matter Particle Physics and CSSM, Department of Physics, University of Adelaide, SA 5005, Australia}

\begin{abstract}
We investigate the interactions of weakly interacting massive particles (WIMPS) with nucleons in nuclear medium by taking into account the effect of nuclear dynamics. 
We derive the nonrelativistic effective operators starting from the relativistic mean-field approximation. 
Certain interactions receive non-negligible corrections, which may significantly change the sensitivities of the WIMP-nucleus scattering cross section to these effective operators.
\end{abstract}

\date{\today}
\maketitle

\section{Introduction}
\label{sec:intro}
Understanding the nature of dark matter (DM) is one of the most important challenges facing modern physics and astronomy. While its existence has been confirmed in a multitude of 
ways~\cite{Ade:2015xua,Abazajian:2014fta}, its nature remains a complete mystery. Initial excitement, motivated by the concept of supersymmetry (SUSY)~\cite{Fayet:1976cr,Ellwanger:2009wj}, about the so-called ``WIMP (weakly interacting massive particle) miracle''~\cite{Bertone:2004pz,Jungman:1995df,Goodman:1984dc} has cooled a little as direct searches around the world have placed ever more stringent limits on the mass and couplings of such 
particles~\cite{Aprile:2017aty,Wang:2019wwo,DEAP:2020,Bondarenko:2019vrb,Kozlov:2019lik}. Nevertheless, whether or not their origin lies in SUSY, WIMPs remain one of the most promising dark matter candidates and it is crucial to pursue any avenue that might provide a significant constraint on the nature of these particles. Such tests range all the way from the effect of captured WIMPs on the properties of neutron stars~\cite{Ellis:2018bkr,Motta:2018rxp,Bertoni:2013bsa,Bell:2013xk} to remarkable experiments designed to directly observe their interactions with 
matter~\cite{Aprile:2017aty,Barak:2020fql}. The latter will be our focus here.

Direct detection searches for WIMP are designed to observe the nuclear recoil events caused by their weak interactions with nuclei in detectors~\cite{Goodman:1984dc, Drukier:1984, Drukier:1986} located deep underground to reduce backgroungs from cosmic rays.
Annual modulation of the count rate in direct detection experiments is a powerful signature for dark matter~\cite{Freese:2012xd}. While such modulation has been observed by the DAMA/LIBRA Collaboration~\cite{DAMA:2008, DAMA/LIBRA:2010,DAMA/LIBRA-2}, other experiments have given null results. 
In a tantalizing hint of a signal, the Xenon1T Collaboration recently reported a $3.5\sigma$ excess of electron recoil event~\cite{Aprile:2020tmw}. 
However, the PandaX-II Collaboration has observed similar event rate, which is also consistent with background-only hypothesis~\cite{PANDAX-II:2020}.
Recent reviews of the theoretical and experimental status  of direct detection can be found in Refs.~\cite{Freese:2012xd}. 

In earth-bound experiments the WIMPs are expected to hit a nuclear target with a velocity of a few hundred kilometres per second, or $\sim 10^{-3} c$. As a consequence, the interaction with a nuclear target necessarily involves low momentum transfer. Both because of this relatively low momentum transfer and the fact that the most sophisticated treaments of nuclear structure tend to be non-relativistic, the theoretical treatment of WIMP-nucleon interactions tends to involve nonrelativistic effective field theory. 
The effective operators were constructed for dark matter of spin 0 and 1/2~\cite{Fan:2010, Fitzpatrick:2013,Anand:2013yka}. 

In such an approach the relativistic Lagrangian density describing the interaction between WIMPs and quarks and gluons is replaced by the most general set of Galilean invariant operators for WIMP-nucleon interactions, including the standard spin-independent and spin-dependent ones. These are obtained by nonrelativistic reduction of the relativistic interactions for a free nucleon. After embedding these operators into nuclei, the WIMP-nucleus cross section can be written in terms of six independent nuclear response functions\cite{Anand:2013yka}. This framework has been widely used in phenomenological analyses of direct detection data~\cite{Zurek:2014, XENON:2017, Kang:2019}.
 
While the most familiar treatments of nuclear structure are non-relativistic, the underlying theory must, of course, be relativistic. This may become important when we consider interactions of a WIMP with a nucleon bound in a nucleus, because the nuclear dynamics, notably the relativistic mean-fields, may modify the effective operators which are to be sandwiched between appropriate nonrelativistic wave functions~\cite{Bolsterli:1974ct}. In particular, as observed in the context of pion production on nuclei, one must take special care with certain interactions when the leading contribution from a relativistic coupling involves the small component of the nucleon 
wavefunction~\cite{Noble:1975,Noble:1979,Bolsterli:1974ct}.
Here we pay special attention to the nonrelativistic reduction of elastic WIMP-nucleon interactions in a nuclear medium in which there are strong Lorentz scalar and vector potentials. 
This analysis can also be applied to the case of arbitrary dark matter spin~\cite{Gondolo:2020}, as well as inelastic scatterings~\cite{Barello:2014}.

In Sec.~\ref{sec:medium-wf}, we show the effect of nuclear dynamics on the nucleon wave functions. 
The modified effective operators in nonrelativistic limit are given in Sec.~\ref{sec:EFT-operators}. We conclude with some remarks on the potential significance of the results presented here in Sec.~\ref{sec:conclusion}.
  

\section{Nucleon wavefunction in medium}
\label{sec:medium-wf}
The relativistic theory of nucleons interacting via the exchange of scalar and vector bosons in the mean-field approximation was initiated by Walecka~\cite{Walecka:1974} and has since been developed extensively~\cite{Horowitz:1981xw,Anastasio:1984gy,Furnstahl:1987fe}. In a nuclear medium, we assume that the wave function of a nucleon bound by Lorentz scalar and vector mean fields satisfies the relativistic Dirac equation
\begin{equation}
\label{eq:Dirac-eq}
\gamma^0 \Big[ - i \gv \cdot \nabla + m_N + V_s + \gamma^0 V_v \Big] \psi(\xv) = E \psi(\xv) \, ,
\end{equation}
where  $V_s$ corresponds to an attractive Lorentz scalar potential and $V_v$ the repulsive fourth component of a four-vector. Phenomenologically these potentials are usually described by the exchange of $\sigma$ and $\omega$ mesons, respectively. Although model dependent, there is a consensus that these potentials must be large (i.e., as much as 30-40\% of the mass of the nucleon) and of opposite sign~\cite{Reinhard:1989zi,Saito:2005rv}. 

We write 
\begin{equation}
\psi = \left( \begin{array}{c}
                              u \\
                              v 
                         \end{array}
                \right) \, , 
\end{equation}
where $u$ and $v$ are the large and small two-component wave functions, respectively, which satisfy
\begin{eqnarray}
\sigmav \cdot \pv v + (m_N + V_s + V_v) u &=& E u \, ,\\
\sigmav \cdot \pv u - (m_N + V_s - V_v) v &=& E v \, .
\end{eqnarray}
Using the above equations, we find 
\begin{eqnarray}
\label{eq:small-v}
v &=& \frac{1}{E + m_N + V_s - V_v} \sigmav \cdot \pv u  
= \frac{1}{2 \widetilde{m}_N} \sigmav \cdot \pv u \, ,
\end{eqnarray}
where 
\begin{equation}
\label{eq:mN-tilde}
\widetilde{m}_N = m_N [1 + ( B + V_s - V_v)/2 m_N]
\end{equation}
with $B = E - m_N$ the binding energy ($B<0$). 
The small component of bound nucleon wave function receives a large correction compared with the free nucleon case, because of the appearance of the combination $(V_s - V_v)$, since at nuclear matter density this is typically larger than half of the mass of the nucleon. 

After eliminating the small component by the replacement in Eq.~(\ref{eq:small-v}), the large component wave function $u(\xv)$ satisfies the non-relativistic Schrodinger equation,
\begin{equation}
\label{eq:pauli-equation-1}
\Big[ \frac{1}{2 \widetilde{m}_N} p^2 + (V_s + V_v) - \frac{1}{2 \widetilde{m}_N^2} \frac{1}{r} \frac{d \widetilde{m}_N}{d r} \sigmav \cdot \lv - \frac{1}{ 4 \widetilde{m}_N^2} \nabla^2 \widetilde{m}_N \Big] u(\xv)= B u(\xv) ,
\end{equation}
which contains a spin-orbit interaction.
If we expand $1/\widetilde{m}_N$ to the first order,
\begin{equation}
\frac{1}{\widetilde{m}_N} = \frac{1}{m_N} \left(1 - \frac{B + V_s - V_v}{2 m_N} \right) ,
\end{equation}
then Eq.~(\ref{eq:pauli-equation-1}) becomes
\begin{eqnarray}
&& \Big[ \frac{1}{2 m_N} p^2 + (V_s + V_v) + \frac{1}{2m_N} (V_s^2 - V_v^2) + \frac{B}{m_N} V_v \nonumber\\
&& - \frac{1}{4 m_N^{2}} \frac{1}{r} \frac{d(V_s - V_v)}{d r} \sigmav \cdot \lv - \frac{1}{ 8 m_N^{2}} \nabla^2 (V_s - V_v) \Big] u(\xv)
= B u(\xv) ,
\end{eqnarray}
from which one can define the effective potential~{\cite{Noble:1978}}
\begin{equation}
V_{\rm eff} = V_s + V_v + \frac{V_s^2 - V_v^2}{2m_N} + \frac{B}{m_N} V_v .
\end{equation}
%


\section{Nonrelativistic WIMP-nucleon operators in medium}
\label{sec:EFT-operators}
In the case of elastic scattering of a WIMP with mass $m_{\chi}$ from a nucleon with mass $m_N$,
we take the incoming (outgoing) momentum of $\chi$ to be $p$ ($p'$) and of $N$ to be $k$ ($k'$).
Galilean invariant combinations of momentum are those made from the momentum transfer $\qv$ and relative incoming velocity $\vv$,
\begin{equation}
\qv = \pv' - \pv, \ \ \ \ \vv = \vv_{\chi, {\rm in}} - \vv_{N, {\rm in}},
\end{equation}
It is common to introduce the related quantity
\begin{equation}
\label{eq:v-perp}
\vv^{\perp} = \vv + \frac{\qv}{2\mu_{\chi N}} = \frac{1}{2} \left( \frac{\pv}{m_{\chi}} + \frac{\pv'}{m_{\chi}} - \frac{\kv}{m_N} - \frac{\kv'}{m_N} \right) \, ,
\end{equation}
which satisfies $\vv^{\perp} \cdot \qv = 0$ by the energy-conservation condition. $\mu_{\chi N}$ is the WIMP-nucleon reduced mass.
\begin{table}[htbp]
\begin{center}
\hspace*{-0.7cm}
\scalebox{0.75}{
\begin{tabular}{cccc}
\hline\hline
$j$  &                         ${\CL}^j_\mathrm{int}$                    &    \multicolumn{2}{c}{Non-relativistic Reduction in medium ($u^{\dag} \CO_{eff} u$) }               \\ 
\hline
1     &  $ \bar{\chi} \chi \bar{N} N$                                      &                                        $ 1_{\chi} 1_N$                                                       &          $ 1_{\chi} 1_N$                               \\[0.4cm]
2     &  $ i \bar{\chi} \chi \bar{N} \gamma^5 N$                  & $ i \frac{\qv}{m_N} \cdot \sv_N \frac{1}{ 1+  (V_s - V_v)/ 2 m_N }$               &           $ i \frac{\qv}{m_N^*} \cdot \sv_N $  \\[0.4cm]
3     &  $ i \bar{\chi} \gamma^5 \chi \bar{N} N$                  & $ - i \frac{\qv}{m_\chi} \cdot \sv_\chi$                                                             &           $ - i \frac{\qv}{m_\chi} \cdot \sv_\chi$     \\[0.4cm]
4     &  $ \bar{\chi} \gamma^5 \chi \bar{N} \gamma^5 N $ & $ - ( \frac{\qv}{m_{\chi}} \cdot \sv_\chi ) ( \frac{\qv}{m_N} \cdot \sv_N ) \frac{1}{ 1+ (V_s - V_v)/ 2 m_N }$         &      $ - ( \frac{\qv}{m_{\chi}} \cdot \sv_\chi ) ( \frac{\qv}{m_N^*} \cdot \sv_N ) $       \\ [0.4cm]
5     &  $\bar{\chi} \gamma^{\mu} \chi \bar{N} \gamma_{\mu} N $   &      $1_{\chi} 1_N$                      &          $1_{\chi} 1_N$           \\ [0.4cm]
6     &  $\bar{\chi} \gamma^{\mu} \chi \bar{N} i \sigma_{\mu\alpha} \frac{q^{\alpha}}{m_M} N$  &  $ \frac{\qv^2}{2 m_N m_M} 1_{\chi} 1_{N}  + 2 ( \frac{\qv}{m_{\chi}} \times \sv_{\chi} + i \vv^{\perp}) \cdot (\frac{\qv}{m_M} \times \sv_N)$
  &      $ \frac{\qv^2}{2 m_N^* m_M} 1_{\chi} 1_{N}  + 2 ( \frac{\qv}{m_{\chi}} \times \sv_{\chi} + i \vv^{*\perp}) \cdot (\frac{\qv}{m_M} \times \sv_N)$      \\ [0.3cm]
\      &           \           &   $ - [ \frac{\qv^2}{2 m_N m_M} 1_{\chi} 1_{N} - 2 i \frac{\kv' + \kv}{2 m_N} \cdot (\frac{\qv}{m_M} \times \sv_N ) ] \frac{V_s - V_v}{2 m_N [1 + (V_s - V_v)/2m_N]}$      &       \           \\ [0.4cm]
7     &  $\bar{\chi} \gamma^{\mu} \chi \bar{N} \gamma_\mu \gamma^5 N$  & $ -2 \vec{S}_N \cdot \vec{v}^\perp +  2 i \sv_{\chi} \cdot (\sv_N \times \frac{\qv}{m_{\chi}}) 
 -  \sv_N \cdot \frac{\kv' + \kv}{m_N} \frac{V_s - V_v}{2 m_N [1 + (V_s - V_v)/2 m_N]}$       &      $ -2 \vec{S}_N \cdot \vv^{*\perp} +  2 i \sv_{\chi} \cdot (\sv_N \times \frac{\qv}{m_{\chi}}) $     \\ [0.4cm]
8     &  $\bar{\chi} i \gamma^{\mu} \chi \bar{N} \sigma_{\mu\alpha} \frac{q^{\alpha}}{m_M} \gamma^5 N$   &     $ 2 i \frac{\qv}{m_M} \cdot \sv_N$               &       $ 2 i \frac{\qv}{m_M} \cdot \sv_N$  \\[0.4cm]
9     & $\bar{\chi} i \sigma^{\mu\nu} \frac{q_{\nu}}{m_M} \chi \bar{N} \gamma_{\mu} N$  &   $- \frac{\qv^2}{2 m_{\chi} m_M} 1_{\chi} 1_{N} - 2 (\frac{\qv}{m_M} \times \sv_{\chi}) \cdot (\frac{\qv}{m_N} \times \sv_N + i \vv^{\perp})$ 
&      $- \frac{\qv^2}{2 m_{\chi} m_M} 1_{\chi} 1_{N} - 2 (\frac{\qv}{m_M} \times \sv_{\chi}) \cdot (\frac{\qv}{m_N^*} \times \sv_N + i \vv^{*\perp})$         \\ [0.3cm]
 \     &                                   \             & $ + 2 (\frac{\qv}{m_M} \times \sv_{\chi}) \cdot ( \frac{\qv}{m_N} \times \sv_N - i \frac{\kv' + \kv}{2 m_N} ) \frac{V_s - V_v}{2 m_N [1 + (V_s - V_v)/2 m_N]} $       &     \     \\ [0.4cm]
10   &  $\bar{\chi} i \sigma^{\mu\nu} \frac{q_{\nu}}{m_M} \chi \bar{N} i \sigma_{\mu\alpha} \frac{q^{\alpha}}{m_M} N$   &     $4 (\frac{\qv}{m_M} \times \sv_{\chi}) \cdot (\frac{\qv}{m_M} \times \sv_N)$       &      $4 (\frac{\qv}{m_M} \times \sv_{\chi}) \cdot (\frac{\qv}{m_M} \times \sv_N)$   \\[0.4cm]
11  &  $\bar{\chi} i \sigma^{\mu\nu} \frac{q_{\nu}}{m_M} \chi \bar{N} \gamma_{\mu} \gamma_5 N$  &  $4 i (\frac{\qv}{m_M} \times \sv_{\chi}) \cdot \sv_N$      &      $4 i (\frac{\qv}{m_M} \times \sv_{\chi}) \cdot \sv_N$  \\[0.4cm]
12  & $ i \bar{\chi} i \sigma^{\mu\nu} \frac{q_{\nu}}{m_M} \chi  \bar{N} i \sigma_{\mu\alpha} \frac{q^{\alpha}}{m_M} \gamma^5 N$   
&   $- [i \frac{\qv^2}{m_{\chi} m_M} - 4 \vv^{\perp} \cdot (\frac{\qv}{m_M} \times \sv_{\chi})] (\frac{\qv}{m_M} \cdot \sv_N) $         
&     $- [i \frac{\qv^2}{m_{\chi} m_M} - 4 \vv^{*\perp} \cdot (\frac{\qv}{m_M} \times \sv_{\chi})] (\frac{\qv}{m_M} \cdot \sv_N) $     \\[0.3cm]      
 \      &     \     &    $+ 4 \frac{\kv' + \kv}{2m_N} \cdot ( \frac{\qv}{m_M} \times \sv_{\chi} ) (\frac{\qv}{m_M} \cdot \sv_N)  \frac{V_s - V_v}{2m_N [1 + (V_s - V_v)/2m_N]}$    &            \          \\[0.4cm] 
13   &  $\bar{\chi} \gamma^\mu \gamma^5 \chi \bar{N} \gamma_{\mu} N $ &    $2 \vec{S}_{\chi} \cdot \vv^{\perp} + 2 i \vec{S}_{\chi} \cdot (\vec{S}_N \times \frac{\qv}{m_N}) $   
      &         $2 \vec{S}_{\chi} \cdot \vv^{*\perp} + 2 i \vec{S}_{\chi} \cdot (\vec{S}_N \times \frac{\qv}{m_N^*}) $  \\[0.3cm]
  \   &    \     &    $+  [ \sv_{\chi} \cdot \frac{\kv' + \kv}{m_N} - 2 i \sv_{\chi} \cdot (\sv_N \times \frac{\qv}{m_N})] \frac{V_s - V_v}{2 m_N [ 1 + (V_s - V _v)/2m_N]}$  &  \     \\ [0.4cm]
14   & $\bar{\chi} \gamma^{\mu}\gamma^5 \chi \bar{N} i \sigma_{\mu\alpha} \frac{q^{\alpha}}{m_M} N$   &  $4 i \sv_{\chi} \cdot (\frac{\qv}{m_M} \times \sv_N) $     &       $4 i \sv_{\chi} \cdot (\frac{\qv}{m_M} \times \sv_N) $    \\[0.4cm]
15   &  $\bar{\chi} \gamma^\mu \gamma^5 \chi \bar{N} \gamma^\mu \gamma^5 N$    &  $-4 \vec{S}_\chi \cdot \vec{S}_N$       &    $-4 \vec{S}_\chi \cdot \vec{S}_N$     \\ [0.4cm] 
16   &  $ i \bar{\chi} \gamma^{\mu} \gamma^5 \chi \bar{N} i \sigma_{\mu\alpha} \frac{q^{\alpha}}{m_M} \gamma^5 N$   &  
 $4 i ( \sv_{\chi} \cdot \vv^{\perp} ) ( \sv_N \cdot \frac{\qv}{m_M} ) + 4 i (\sv_{\chi} \cdot \frac{\kv' + \kv}{2m_N} ) ( \sv_N \cdot \frac{\qv}{m_M} ) \frac{V_s - V_v}{2 m_N [ 1 + (V_s - V_v)/2m_N]}$    &    $4 i ( \sv_{\chi} \cdot \vv^{*\perp} ) ( \sv_N \cdot \frac{\qv}{m_M} ) $ \\[0.4cm] 
17   &  $ i \bar{\chi} i \sigma^{\mu\nu} \frac{q_{\nu}}{m_M} \gamma^5 \chi \bar{N} \gamma_{\mu} N$   &     $2 i \frac{\qv}{m_M} \cdot \sv_{\chi}$      &   $2 i \frac{\qv}{m_M} \cdot \sv_{\chi}$   \\[0.4cm] 
18   &  $ i \bar{\chi} i \sigma^{\mu\nu} \frac{q_{\nu}}{m_M} \gamma^5 \chi \bar{N} i \sigma_{\mu\alpha} \frac{q^{\alpha}}{m_M} N$   &   $( \frac{\qv}{m_M} \cdot \sv_{\chi} ) [ i \frac{\qv^2}{m_N m_M} - 4 \vv^{\perp} \cdot (\frac{\qv}{m_M} \times \sv_N)] $       &      $( \frac{\qv}{m_M} \cdot \sv_{\chi} ) [ i \frac{\qv^2}{m_N^* m_M} - 4 \vv^{*\perp} \cdot (\frac{\qv}{m_M} \times \sv_N)] $     \\[0.3cm]
\     &   \    &   $- ( \frac{\qv}{m_M} \cdot \sv_{\chi} ) [ i \frac{\qv^2}{m_N m_M} + 4 \frac{\kv' + \kv}{2 m_N} \cdot (\frac{\qv}{m_M} \times \sv_{N})] \frac{V_s - V_v}{2 m_N [ 1 + (V_s - V_v)/2m_N]}$    &     \     \\[0.4cm]
19   &  $ i \bar{\chi} i \sigma^{\mu\nu} \frac{q_{\nu}}{m_M} \gamma^5 \chi \bar{N} \gamma_{\mu} \gamma^5 N$   &     $- 4 i ( \sv_{\chi} \cdot \frac{\qv}{m_M} ) ( \sv_N \cdot \vv^{\perp} ) - 4 i ( \sv_{\chi} \cdot \frac{\qv}{m_M} ) ( \sv_N \cdot \frac{\kv' + \kv}{2 m_N} ) \frac{V_s - V_v}{2m_N[1 + (V_s - V_v)/2m_N]}$     &     $- 4 i ( \sv_{\chi} \cdot \frac{\qv}{m_M} ) ( \sv_N \cdot \vv^{*\perp} ) $  \\[0.4cm]
20   &  $ i \bar{\chi} i \sigma^{\mu\nu} \frac{q_{\nu}}{m_M} \gamma^5 \chi \bar{N} i \sigma_{\mu\alpha} \frac{q^{\alpha}}{m_M} \gamma^5 N$   &     $4 i ( \frac{\qv}{m_M} \cdot \sv_{\chi } ) ( \frac{\qv}{m_M} \cdot \sv_{N} )$    &   $4 i ( \frac{\qv}{m_M} \cdot \sv_{\chi } ) ( \frac{\qv}{m_M} \cdot \sv_{N} )$  \\[0.4cm]
\hline
\end{tabular}
}
\caption{ {\small The relativistic interactions $\CL^j_\mathrm{int}$ are listed in the second column. The corresponding effective operators 
after non-relativistic reduction are given in the third and fourth column. }}
 \label{table:operators}
\end{center}
\end{table}

Following the procedure in Ref.~\cite{Anand:2013yka}, we rederive the nonrelativistic reductions of the various relativistic interactions.
To leading order in $p/m_{\chi}$ and $k/m_N$, the effective operators 
can be obtained from those for a free nucleon, with $m_N$ being replaced by $\widetilde{m}_N$.

As an approximation, we can simply neglect the binding energy $B$ in Eq.~(\ref{eq:mN-tilde}). 
The explicit form of the effective operators are listed in the third column of Tab.~\ref{table:operators}, 
with the corrections being separated from the free nucleon results.
The nonrelativistic analog of invariant amplitudes can be obtained from the matrix elements of these operators between the wave functions $u^{\dag}$ and $u$.
In the absence of the potentials, these operators reduce to those for a free nucleon given in 
Ref.~\cite{Anand:2013yka}.

Another useful approximation is  to rewrite Eq.~(\ref{eq:mN-tilde}) as
\begin{equation}
\widetilde{m}_N = m_N^* + \frac{B- (V_s + V_v)}{2} ,
\end{equation}
where $m_N^* = m_N + V_s$ is the nucleon effective mass commonly used in nuclear dynamics. 
By neglecting the small quantity $B - (V_s + V_v)$,
the corresponding nonrelativistic operators are shown in the fourth column of Tab.~\ref{table:operators}, 
where $\vv^{*\perp}$ is defined by Eq.~(\ref{eq:v-perp}) with $m_N \rightarrow m_N^*$.

Taking the vector-tensor coupling, case 6, as an example, we see that the correction induced by the relativistic mean-field potentials typically involves a factor $1/[1+(V_s - V_v)/2m_N]$. 
As the scalar interaction is attractive and the vector interaction repulsive, this results in a boost for the effective interaction.  
In quantum hadrodynamics (QHD), the typical values of the potentials are $V_s \approx - 400~{\rm MeV}$ and $V_v \approx 350~{\rm MeV}$~\cite{QHD:1997}, respectively,  
while the quark meson coupling (QMC) model gives smaller results, $V_s \approx - 190~{\rm MeV}$ and $V_v \approx 130~{\rm MeV}$~\cite{Saito:2005rv}.
As a result, the corrections to some of these interactions will be as large as 21-66\%, 
which may significantly change the sensitivities of WIMP-nucleus scattering cross section to the effective operators, compared with the free nucleon results.

\section{Conclusion}
\label{sec:conclusion}
The existence of strong Lorentz scalar and vector mean fields in atomic nuclei leads to corrections to some of the nonrelativistic effective interactions which determine the cross section for dark matter scattering. This is only potentially important when the combination $(V_s - V_v)/2m_N$ makes an appearance and then it can lead to a correction as large as 21-66\%, depending where in the nucleus the interaction takes place. In a nucleus with $N \neq Z$, the isovector interaction has a Lorentz vector from which will increase the strength of the interaction with neutrons and decrease it for protons. This may well be significant in relating the results of experiments on different nuclear targets once  a dark matter signal is found.

\section*{Acknowledgements}
This work was supported by the University of Adelaide and the Australian Research Council through the Centre of Excellence for Dark Matter Particle Physics (CE200100008) and Discovery Project DP180100497.



\begin{thebibliography}{99}
%
\bibitem{Ade:2015xua}
P.~A.~R.~Ade \textit{et al.} [Planck],
Astron. Astrophys. \textbf{594}, A13 (2016).
%
\bibitem{Abazajian:2014fta}
K.~N.~Abazajian, N.~Canac, S.~Horiuchi and M.~Kaplinghat,
Phys. Rev. D \textbf{90}, 023526 (2014).
%
\bibitem{Fayet:1976cr}
P.~Fayet and S.~Ferrara,
Phys. Rept. \textbf{32}, 249 (1977).
%
\bibitem{Ellwanger:2009wj}
U.~Ellwanger,
Nucl. Phys. B Proc. Suppl. \textbf{200-202}, 113 (2010).
%

\bibitem{Bertone:2004pz}
G.~Bertone, D.~Hooper and J.~Silk,
Phys. Rept. \textbf{405}, 279 (2005).
%

\bibitem{Jungman:1995df}
G.~Jungman, M.~Kamionkowski and K.~Griest,
Phys. Rept. \textbf{267}, 195 (1996).
%

\bibitem{Goodman:1984dc}
M.~W.~Goodman and E.~Witten,
Phys. Rev. D \textbf{31}, 3059 (1985).
%

\bibitem{Aprile:2017aty}
E.~Aprile \textit{et al.} (XENON Collaboration),
Eur. Phys. J. C \textbf{77}, 881 (2017).
%
\bibitem{Wang:2019wwo}
Y.~Wang \textit{et al.} (CDEX Collaboration),
Phys. Rev. D \textbf{101}, 052003 (2020).
%


\bibitem{DEAP:2020}
P.~Adhikari {\it et al.} (DEAP Collaboration), Phys. Rev. D {\bf 102}, 082001 (2020).


\bibitem{Bondarenko:2019vrb}
K.~Bondarenko, A.~Boyarsky, T.~Bringmann, M.~Hufnagel, K.~Schmidt-Hoberg and A.~Sokolenko,
J. High Energy Phys. \textbf{03}, 118 (2020).
%

\bibitem{Kozlov:2019lik}
A.~Kozlov, D.~Chernyak, Y.~Takemoto, K.~Fushimi, K.~Imagawa, K.~Yasuda, H.~Ejiri, R.~Hazama, H.~Ikeda and K.~Inoue, \textit{et al.}
J. Phys. Conf. Ser. \textbf{1390}, 012118 (2019).
%

\bibitem{Ellis:2018bkr}
J.~Ellis, G.~H\"utsi, K.~Kannike, L.~Marzola, M.~Raidal and V.~Vaskonen,
Phys. Rev. D \textbf{97}, 123007 (2018).
%
\bibitem{Motta:2018rxp}
T.~F.~Motta, P.~A.~M.~Guichon and A.~W.~Thomas,
J. Phys. G \textbf{45}, 05LT01 (2018).
%
\bibitem{Bertoni:2013bsa}
B.~Bertoni, A.~E.~Nelson and S.~Reddy,
Phys. Rev. D \textbf{88}, 123505 (2013).
%
\bibitem{Bell:2013xk}
N.~F.~Bell, A.~Melatos and K.~Petraki,
Phys. Rev. D \textbf{87}, 123507 (2013).
%
\bibitem{Barak:2020fql}
L.~Barak \textit{et al.} (SENSEI Collabration),
Phys. Rev. Lett. \textbf{125}, 171802 (2020).
%
\bibitem{Drukier:1984}
A.~Drukier, and L.~Stodolsky, Phys. Rev. D {\bf 30}, 2295 (1984).


\bibitem{Drukier:1986}
A.~Drukier, K.~Freese, and D.~Spergel, Phys. Rev. D {\bf 33}, 3495(1986).
%
\bibitem{Freese:2012xd}
K.~Freese, M.~Lisanti and C.~Savage,
Rev. Mod. Phys. \textbf{85}, 1561 (2013).
%
\bibitem{DAMA:2008}
R.~Bernabei {\it et al.} (DAMA Collaboration), Eur. Phys. J. C {\bf 56}, 333 (2008).

\bibitem{DAMA/LIBRA:2010}
R.~Bernabei {\it et al.} (DAMA and LIBRA Collaborations), Eur. Phys. J. C {\bf 67}, 39 (2010).

%

\bibitem{DAMA/LIBRA-2}
R.~Bernabei {\it et al.}, Universe {\bf 4}, 116 (2018).


\bibitem{Aprile:2020tmw}
E.~Aprile \textit{et al.} (XENON Collaboration),
Phys. Rev. D \textbf{102}, 072004 (2020)


\bibitem{PANDAX-II:2020}
X.~Zhou {\it et al.} (PandaX-II Collaboration), Chin. Phys. Lett. {\bf 38}, 011301 (2021).


\bibitem{Fan:2010}
J.~Fan, M.~Reece, and L.-T.~Wang, J. Cosmol. Astropart. Phys. {\bf 11} (2010) 042.

\bibitem{Fitzpatrick:2013}
A.~L.~Fitzpatrick, W.~C.~Haxton, E.~Katz, N.~Lubbers, and Y.~Xu, J. Cosmol. Astropart. Phys. {\bf 02} (2013) 004.

%
\bibitem{Anand:2013yka}
N.~Anand, A.~L.~Fitzpatrick and W.~C.~Haxton,
Phys. Rev. C \textbf{89}, 065501 (2014)
%


\bibitem{Zurek:2014}
M.~I.~Gresham, and K.~M.~Zurek, Phys. Rev. D {\bf 89}, 123521 (2014).

\bibitem{XENON:2017}
E.~Aprile {\it et al.} (XENON Collaboration), Phys. Rev. D {\bf 96}, 042004 (2017).

\bibitem{Kang:2019}
S.~Kang, S.~Scopel, and G.~Tomar, Phys. Rev. D {\bf 99}, 103019 (2019).
%


\bibitem{Bolsterli:1974ct}
M.~Bolsterli, W.~R.~Gibbs, B.~F.~Gibson and G.~J.~Stephenson,
Phys. Rev. C \textbf{10}, 1225 (1974).
%

\bibitem{Noble:1975}
J.~M.~Eisenberg, J.~V.~Noble, and H.~J.~Weber, Phys. Rev. C {\bf 11}, 1048 (1975).

\bibitem{Noble:1979}
J.~V.~Noble, Phys. Rev. Lett. {\bf 43}, 100 (1979).


\bibitem{Gondolo:2020}
P.~Gondolo, S.~Kang, and G.~Tomar, arXiv: 2008.05120 [hep-ph].

\bibitem{Barello:2014}
G.~Barello, S.~Chang, and C.~A.~Newby, Phys. Rev. D {\bf 90}, 094027 (2014).


\bibitem{Walecka:1974}
J.~D.~Walecka, Ann. Phys. (NY) {\bf 83}, 491 (1974).
%

\bibitem{Horowitz:1981xw}
C.~J.~Horowitz and B.~D.~Serot,
Nucl. Phys. A \textbf{368}, 503 (1981).
%

\bibitem{Anastasio:1984gy}
M.~R.~Anastasio, L.~S.~Celenza, W.~S.~Pong and C.~M.~Shakin,
Phys. Rept. \textbf{100}, 327 (1983).
%

\bibitem{Furnstahl:1987fe}
R.~J.~Furnstahl, C.~E.~Price and G.~E.~Walker,
Phys. Rev. C \textbf{36}, 2590 (1987).
%

\bibitem{Reinhard:1989zi}
P.~G.~Reinhard,
Rept. Prog. Phys. \textbf{52}, 439 (1989).
%

\bibitem{Saito:2005rv}
K.~Saito, K.~Tsushima and A.~W.~Thomas,
Prog. Part. Nucl. Phys. \textbf{58}, 1 (2007).
%


\bibitem{Noble:1978}
J.~V.~Noble, Phys. Rev. C {\bf 17}, 2151 (1978).
%

\bibitem{QHD:1997}
B.~D.~Serot, J.~D.~Walecka, Int. J. Mod. Phys. E {\bf 6}, 515 (1997).




\end{thebibliography}
\end{document}